\documentclass{article}
\usepackage{amssymb}
\usepackage{graphicx}
\usepackage{tikz}
\usepackage{amsmath}
\usepackage{subcaption}
\usepackage{booktabs}
\usepackage{caption}
\usepackage{todonotes}
\usepackage{multirow}
\usepackage{amsfonts}
\usepackage{rotating}
\usepackage{url}
\usepackage{color}
\usepackage{xcolor}
\usepackage{array}
\usepackage{mathtools}
\usepackage{comment}
\usepackage{dblfloatfix}
\usepackage{enumitem}
\usepackage{hyperref}
\usepackage{algorithm}
\usepackage{algpseudocode}
\usepackage[margin=1.1in]{geometry}
\usepackage{multirow}
\usepackage{amsthm}
\usepackage{amsmath}
\usepackage{algorithmicx}
\usepackage[official]{eurosym}
\usepackage{pgfplots}
\usepackage{tablefootnote}

\usepackage{natbib}

\newtheorem{defi}{Definition}
\graphicspath{{Graphics/}}
\newcolumntype{L}[1]{>{\raggedright\arraybackslash}p{#1}}

\usepackage[affil-it]{authblk}

\newtheorem{theo}[defi]{Theorem}
\newtheorem{lem}[defi]{Lemma}

\title{An Approximation Algorithm for Multi Allocation Hub Location Problems}
\author{Niklas Jost}

\date{}
\begin{document}
\maketitle

\begin{abstract}
	The multi allocation $p$-hub median problem (MApHM), the multi allocation uncapacitated hub location problem (MAuHLP) and the multi allocation $p$-hub location problem (MApHLP) are common hub location problems with several practical applications. HLPs aim to construct a network for routing tasks between different locations. Specifically, a set of hubs must be chosen and each routing must be performed using one or two hubs as stopovers. The costs between two hubs are discounted by a parameter $\alpha$. The objective is to minimize the total transportation cost in the MApHM and additionally to minimize the set-up costs for the hubs in the MAuHLP and MApHLP. In this paper, an approximation algorithm to solve these problems is developed, which improves the approximation bound for MApHM to $3.451$, for MAuHLP to $2.173$ and for MApHLP to $4.552$ when combined with the algorithm of \cite{benedito2019approximation}. 
	
	The proposed algorithm is capable of solving much bigger instances than any exact algorithm in the literature. New benchmark instances have been created and published for evaluation, such that HLP algorithms can be tested and compared on huge instances. The proposed algorithm performs on most instances better than the algorithm of \cite{benedito2019approximation}, which was the only known approximation algorithm for these problems by now.
	
\end{abstract}

\section{Introduction}

Hub location problems (HLPs) frequently appear for logistics service providers. They must decide where to open depots such that different locations are connected as efficiently as possible. Often tours start with a pre-carriage milk run to collect multiple parcels in an area to bring them to a local depot or branch. These parcels are delivered to the destination branch in the main carriage. In the on-carriage, the parcels are again delivered by a milk run.

In this paper, the focus is on optimizing the main carriage step. Instead of direct transports between any pair of branches, the parcels are delivered to central warehouses, transshipment points or hubs in between. This has two main benefits: First, many parcels can be transported together, although they have different destinations. This results in consolidation effects such as lower costs and a better network structure. The second advantage is that the mode of transport can be changed and multimodal transportation can be used, which is also more efficient. 

The task is to identify hubs for building an efficient transport network. The transportation costs between hubs are reduced to model the consolidation benefits of multimodal transportation. In addition, one or two hubs must be chosen as stopovers for any transport.

The hub location model has been introduced by \cite{o1986location}. Later, \cite{campbell1994integer} made integer programming formulations for various HLPs as the $p$-hub median problem (pHM) or the uncapacitated hub location problem (uHLP), which are the most common HLPs. Reviews of HLPs can be found in \cite{alumur2008network} and \cite{campbell2012twenty}. 

Many algorithms were developed for the multi allocation $p$-hub median problem (MApHM) to solve large-scale instances efficiently. A greedy-interchange heuristic was presented by \cite{campbell1996hub}. Two years later, an efficient mathematical formulation was created by \cite{ernst1998exact}. A special case where only one hub can be chosen between origin and destination was considered by \cite{sasaki1999selection}. 

For the multi allocation uncapacitated hub location problem (MAuHLP) branch and bound algorithms were developed by \cite{klincewicz1996dual} and \cite{mayer2002hublocator}. Later, \cite{canovas2007solving} presented a dual-ascent branch and bound heuristic. 

The multi allocation hub location problem (MAHLP) is a combination and therefore a generalization of MApHM and MAuHLP. 

To the best of the authors' knowledge, only \cite{benedito2019approximation} have given an approximation algorithm for these problems. They have already constructed a sophisticated approximation algorithm for the single allocation variants, resulting in a $6.35$ approximation for the SApHM, $2.48$ for the SAuHLP and $8.47$ for SApHLP, such that this paper focuses on the multi allocation variants. For them they have shown a $3.68$ approximation algorithm for the MApHM, $2.49$ for the MAuHLP and $4.74$ for MAHLP.

A typical multi allocation problem is designing a transport network. Hence, this problem appears for any logistics service provider. For establishing a complex network structure with hundreds of branches and potential hubs the discussed exact algorithms will not be able to give a solution in a reasonable time by the complexity of the problem. In this scenario it is crucial to have a faster algorithm, such as the proposed one. Further applications are telecommunication networks, postal companies and the aviation.

In the next section, ILP formulations of the problems are given. The reduction-based algorithm is explained and presented in section \ref{algosec}. In section \ref{nor} the approximation bound is proven. The prove of the used lemmata is outsourced to section \ref{proof}. In the last section (\ref{prac}), the quality of the algorithm's solution is tested on several instances.

\section{Mathematical Model}

\begin{figure}[H]
	\centering
	\begin{tikzpicture}
		\node[shape=circle,draw=lightgray] (b) at (-0.3,0) {$H_3$};
		\node[shape=circle,draw=blue,minimum size=30pt] (a) at (1,0) {$B_2$};
		\node[shape=circle,draw=lightgray] (d) at (5,0) {$H_4$};;
		\node[shape=circle,draw=blue,minimum size=30pt] (c) at (9,0) {$B_3$};
		
		\node[shape=circle,draw=lightgray] (b2) at (10.3,0) {$H_5$};
		\node[shape=circle,draw=lightgray] (a2) at (3.7,3) {$H_1$};
		\node[shape=circle,draw=blue,minimum size=30pt] (d2) at (5,3) {$B_1$};
		\node[shape=circle,draw=lightgray] (c2) at (6.3,3) {$H_2$};
		\path [->, bend right,line width=0.4mm](d2) edge node[left] {} (a2);		
		\path [->, bend left,draw=red,line width=0.4mm](d2) edge node[left] {} (a2);
		\path [->,line width=0.4mm](a2) edge node[left] {} (b);
		\path [->,line width=0.4mm](b) edge node[left] {} (a);
		\path [->,line width=0.4mm](d2) edge node[left] {} (c2);
		\path [->,line width=0.4mm](c2) edge node[left] {} (b2);
		\path [->,line width=0.4mm](b2) edge node[left] {} (c);
		\path [->,draw=red,line width=0.4mm](a2) edge node[left] {} (d);
		\path [->,draw=red,line width=0.4mm](d) edge node[left] {} (a);
		\path [->,draw=red,line width=0.4mm](d) edge node[left] {} (c);
		
		\node[shape=circle,draw=none] (ctz) at (7,0.2) {$3$};
		\node[shape=circle,draw=none] (ctz) at (3,0.2) {$3$};	
		\node[shape=circle,draw=none] (ctz) at (4.8,1.5) {$3$};
		
		\node[shape=circle,draw=none] (ctz) at (1.5,1.6) {$5$};
		\node[shape=circle,draw=none] (ctz) at (7.8,1.6) {$5$};
		\node[shape=circle,draw=none] (ctz) at (0.3,-0.3) {$1$};
		\node[shape=circle,draw=none] (ctz) at (9.75,-0.3) {$1$};
		\node[shape=circle,draw=none] (ctz) at (4.3,3.5) {$1$};
		\node[shape=circle,draw=none] (ctz) at (5.65,3.3) {$1$};	
		
	\end{tikzpicture}
	\caption{A hub location problem and two (black/red) possible solutions} \label{1bild}
\end{figure}
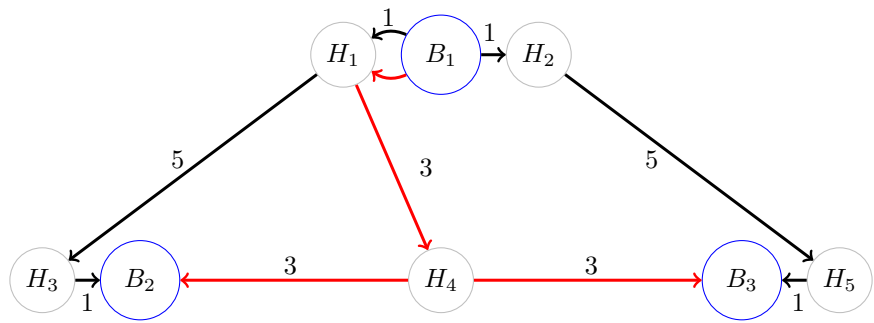

This paper focuses on the multi allocation variant of this strategic, offline problem. Unlike the single allocation variant, each delivery task can be planned individually. In the single allocation variant, each delivery task starting at the same branch must use the same first hub. As an example, consider three branches $\mathcal{B}=\{B_1,B_2,B_3\}$, five hubs $\mathcal{H}=\{H_1,...,H_5\}$ and given delivery tasks $\mathcal{T}=\{(B_1,B_2),(B_1,B_3)\}$. Possible solutions for a given distance function $d$ are illustrated in Figure \ref{1bild}.

In the following, a sequence of an origin branch, one or two hubs and a destination branch is called a tour for that pair of branches. In the multi allocation variant, it could be reasonable to use the tour $B_1\rightarrow H_1 \rightarrow H_3 \rightarrow B_2$ and $B_1\rightarrow H_2 \rightarrow H_5 \rightarrow B_3$ (black arrows). Alternatively, if fewer hubs should be opened, it might also be reasonable to route everything over $H_1$ and $H_4$ instead of $H_3$ or $H_5$ (red arrows). In the single allocation variant, it is not possible to connect $A$ to $H_1$ and $H_2$. Instead, one hub must be used for both deliveries, as in the red solution. Notice that a single allocation solution is always a feasible multi allocation solution. Furthermore, an optimal multi allocation routing can be computed fast for a given set of open hubs by testing any possible combination of hubs for any tour.

In the model, a graph $G=(V,E)$ with edge weights $d_{ij}\in \mathbb{R}_0^+$ for $i,j\in V$, delivery tasks $\mathcal{T}$, a set of branches $\mathcal{B}\subseteq V$ and a set of potential hubs $\mathcal{H}\subseteq V$ is given. Any vertex is a branch, a hub or both such that $V=\mathcal{B} \cup \mathcal{H}$. The decision variable $X_{ijkm}\in \{0,1\}$ indicates if the tour from $i$ to $m$ over $j$ and $k$ with $j,k\in \mathcal{H}$, meaning the tour $i\rightarrow j \rightarrow k \rightarrow m$ is used. As discussed, there are consolidation effects between two hubs, such that using the connection between them is cheaper or faster. This is modeled by a given discount factor $\alpha$ with $0\leq \alpha\leq 1$. The costs for the tour $i\rightarrow j \rightarrow k \rightarrow m$ therefore are \[d_{ij}+\alpha d_{jk}+d_{km}.\] 
In the example above, for a given $\alpha=\frac{1}{2}$, the cost of any red tour is $1+\frac{1}{2}\cdot 3+3=5.5$ and the cost of any black tour is $1+\frac{1}{2}\cdot 5+1=4.5$.

In the next subsection, some restrictions on the distance function are given. For instance, $d_{ii}=0$ will be assumed (\eqref{ident}). Consequently, the case where only one hub is used in a tour can be modelled by setting $j=k$ without having costs for the $\alpha d_{jk}$ term.

Furthermore, the binary decision variable $Y_i$ indicates whether hub $i$ is open, and only open hubs can be used for the routing. Since it is expensive to open hubs, opening them is limited. In the pHM, the number of open hubs is limited by a given integer $p\in \mathbb{N_+}$. The objective is to minimize the summed transportation costs. In the uHLP, the hubs have opening costs $c_{h_1},c_{h_2},...,c_{h_{|\mathcal{H}|}}$. In the pHLP the number of hubs is limited and opening costs need to be considered.

To simplify the notation, let $T_{b,b'}=1$ iff $(b,b')\in \mathcal{T}$ is a given delivery tasks and $T_{b,b'}=0$ else.
The ILP variables are as follows:

\begin{itemize}
	\item $p\in \mathbb{N}$: maximum number of open hubs (for the pHM and pHLP)
	\item $c_{h_1},c_{h_2},...,c_{h_{|\mathcal{H}|}}$: set up costs of the hubs (for the uHLP and pHLP)
	\item $\mathcal{B}$: a finite set of branches
	\item $\mathcal{H}$: a finite set of potential hubs
	\item $d_{i,j}$: a non negative distance function for any $i,j\in \mathcal{B}\cup \mathcal{H}$
	\item $\tau=\{(B_i,B_j),....\}$: a set of delivery tasks 
\end{itemize}
In addition, the following decision variables are used:
\begin{itemize}
	\item $Y_i\in\{0,1 \}$: deciding if hub $i$ is opened
	\item $X_{bhh'b'}\in\{0,1 \}$: deciding if the corresponding tour is applied
\end{itemize}

Then the ILP for the $p$-hub median problem is:

\begin{align} 
	\min& \sum_{(b,b')\in \tau}\sum_{h\in\mathcal{H}}\sum_{h'\in\mathcal{H}} (d_{bh}+\alpha \cdot d_{hh'}+d_{h'b'})\cdot X_{bhh'b'}\\
	\label{med0}s.t.&\sum_{h\in \mathcal{H}}Y_h\leq p&&\\
	\label{med3}&\sum_{h\in \mathcal{H}}\sum_{h'\in \mathcal{H}}X_{bhh'b'}=T_{b,b'}&&\forall b,b'\in \mathcal{B}\\
	\label{med4}&X_{bhh'b'}\leq Y_h && \forall b,b'\in \mathcal{B}; \forall h,h'\in \mathcal{H}\\
	\label{med5}&X_{bhh'b'}\leq Y_{h'}&& \forall b,b'\in \mathcal{B}; \forall h,h'\in \mathcal{H}\\
	\label{med1}&Y_h\in \{0,1\} && \forall h\in \mathcal{H}\\
	\label{med2}&X_{bhh'b'}\in \{0,1\} && \forall b,b'\in \mathcal{B}; \forall h,h'\in \mathcal{H}
\end{align}

Constraint \eqref{med0} ensures that at most $p$ hubs are opened. The existence of exactly one routing for any pair of branches in~$\mathcal{T}$ is ensured by \eqref{med3} and \eqref{med2}. Constraint, \eqref{med4} together with \eqref{med5} restrict tours to use open hubs only. Lastly, by constraint \eqref{med1}, every hub is either open or closed.

In the uncapacitated hub location problem, constraint \eqref{med0} is not applied and the objective function is replaced by \[\min \sum_{(b,b')\in \tau}\sum_{h\in\mathcal{H}}\sum_{h'\in\mathcal{H}} (d_{b,h}+\alpha \cdot d_{hh'}+d_{h'b'})\cdot X_{bhh'b'}+\sum_{h\in \mathcal{H}}Y_h\cdot c_h.\]

As mentioned the $p$-hub location problem is a generalization of both problems, such that any constraint from the pHM together with the objective of the uHLP is applied. Setting $c_f=0$ for any facility would result in the pHM and setting $p=|\mathcal{H}|$ would result in the uHLP.

The problems will be reduced to the corresponding facility location problem (FLP); namely the $k$-median problem, the uncapacitated facility location problem (uFLP) and the $k$-facility location problem ($k$-FLP). Then the solution of a $k$-median/ facility location algorithm will be used as hub location solution. To improve the readability of this paper, when talking about FLPs the $k$-median problem will be meant as well.


\subsection{Facility location problems}

For a given set of cities $\mathcal{C}$ and facilities $\mathcal{F}$, the task is to open facilities and connect any city to exactly one open facility. The objective is to minimize the summed distances (added to the set-up costs for uFLP and $k$-FLP). Different notation is used to clarify when talking about the HLP and when about the FLP.

In the FLPs, the following variables exist:

\begin{itemize}
	\item $k\in \mathbb{N}$: maximum number of open facilities (for the $k$-median and $k$-FLP)
	\item $c_{F_1,F_2,...F_{|\mathcal{F}|}}$: set up costs (for the uFLP and $k$-FLP)
	\item $\mathcal{C}$: a finite set of cities
	\item $\mathcal{F}$: a finite set of potential facilities
	\item $\Gamma_{i,j}$: a non negative distance function for any $i,j\in \mathcal{C}\cup \mathcal{F}$
\end{itemize}

In addition, the following decision variables are used:

\begin{itemize}
	\item $Y_f$: deciding if facility $f$ is opened
	\item $X_{c,f}$: deciding if city $c$ is connected to facility $f$
\end{itemize}

The ILP for the $k$-FLP is:

\begin{align} 
	\min& \sum_{c\in\mathcal{C}}\sum_{f\in\mathcal{F}} \Gamma_{c,f}\cdot X_{c,f}+\sum_{f\in \mathcal{F}}Y_f\cdot c_{f}\\
	\label{hlp0}s.t.&\sum_{f\in \mathcal{F}}Y_f\leq k&&\\
	\label{hlp3}&\sum_{f\in \mathcal{F}}X_{c,f}=1&&\forall c\in \mathcal{C}\\
	\label{hlp4}&X_{c,f}\leq Y_f && \forall c\in \mathcal{C}; \forall f\in \mathcal{F}\\
	\label{hlp1}&Y_f\in \{0,1\} && \forall f\in \mathcal{F}\\
	\label{hlp2}&X_{c,f}\in \{0,1\} && \forall c\in \mathcal{C}; \forall f\in \mathcal{F}.
\end{align}

By constraint \eqref{hlp0} at most $k$ facilities open. Any customer is served by constraint \eqref{hlp3} and only open facilities are used by \eqref{hlp4}. In the uFLP, constraint \eqref{hlp0} is not applied. In the $k$-median problem, the objective function is replaced by \[\min \sum_{c\in\mathcal{C}}\sum_{f\in\mathcal{F}} \Gamma_{c,f}\cdot X_{c,f}.\] 

Notice that an FLP instance can easily be modelled as HLP by adding a city $C'$ and a facility $F'$ with distances $\Gamma_{C',F' }=0$, $\Gamma_{C',F}=\infty$, $\Gamma_{C,F'}=\infty$ for any $F\neq F'$ and $C\neq C'$. Let any city be a branch and any facility a hub. Furthermore, set $\tau= \{(C_1,C'),(C_2,C'),... \}$, $\alpha=0$ and $c_{F'}=0$. Any tour will be connected to $C'$ using $F'$ and the first hub can be interpreted as the facility connected to the corresponding city. Then the $k$-median problem is directly transferred into the $(p+1)$HM, the uFLP into the uHLP and the $k$-FLP into the $(p+1)$HLP.  

For the FLP, many algorithms were established, such as the $2.675+ \epsilon$-approximation algorithm of \cite{byrka2017improved} for the $k$-median problem, the $2+\sqrt{3}+\epsilon$-approximation algorithm of \cite{zhang2007new} for the $k$-FLP and a primal-dual algorithm by \cite{galvao1989method} and a $1.488$ approximation by \cite{li20131} for the uFLP. A survey on FLP can be found by \cite{ulukan2015survey}.

\subsection{Distance function} 
Since most FLP algorithms consider metric distance functions, it needs to be assured that $\Gamma$ forms a metric. 

A nonnegative distance function $ \Gamma$ is metric if the following three conditions hold:

\begin{equation}\label{ident}
	\forall i: \Gamma_{i,i}=0 \qquad \qquad \text{(Definite)}
\end{equation}
\begin{equation}\label{sym}
	\forall i,j: \Gamma_{i,j}=\Gamma_{j,i} \qquad \text{(Symmetry)}
\end{equation}   
\begin{equation}\label{dreieck}
	\forall i,j,k: \Gamma_{i,j}+\Gamma_{j,k}\geq \Gamma_{i,k} \qquad \text{(Triangle inequality)}
\end{equation}

A $p$-norm is a special metric which is defined for two $d$-dimensional points $X=(x_1,x_2,...,x_d)$ and $Y=(y_1,y_2,...,y_d)$ as $\Gamma_{X,Y}=||X-Y||_p=\sqrt[p]{\sum_{i=1}^d |x_i-y_i|^p}$.

\section{The algorithm} \label{algosec}
In this section, a new approximation algorithm for the metric multi allocation $p$-hub median problem (MApHM), the metric multi allocation uncapacitated hub location problem (MAuHLP) and the metric multi allocation $p$-hub location problem (MApHLP) is established. This is done by reducing it to the corresponding FLP, where a $2.675+\epsilon$-approximation algorithm for the $k$-median problem by \cite{byrka2017improved}, a $2+\sqrt{3}+\epsilon$-approximation algorithm for the $k$-FLP by \cite{zhang2007new} and a $1.488$-approximation algorithm for the uFLP by \cite{li20131} exist.\underline{} 

\begin{figure}[H]
	\centering
	\begin{tikzpicture}
		\node[shape=circle,draw=green,minimum size=25pt] (x) at (1.5,2) {$H_1$};
		\node[shape=circle,draw=green,minimum size=25pt] (y) at (11.4,1.3) {$H_2$};
		\node[shape=circle,draw=green,minimum size=25pt] (x2) at (-1.06,0) {$H_{1'}$};
		\node[shape=circle,draw=green,minimum size=25pt] (y2) at (6.6,1.3) {$H_{2'}$};
		\node[shape=circle,draw=blue,minimum size=30pt] (a) at (1,0) {$B_1$};
		\node[shape=circle,draw=blue,minimum size=30pt] (c) at (9,0) {$B_2$};
		
		\path [->,draw=red,line width=0.4mm](x) edge node[left] {} (y2);
		\path [->,draw=green,line width=0.4mm](a) edge node[left] {} (x);
		\path [->,draw=green,line width=0.4mm](y) edge node[left] {} (c);
		\path [->,draw=green,line width=0.4mm](a) edge node[left] {} (x2);
		\path [->,draw=green,line width=0.4mm](y2) edge node[left] {} (c);
		
	\end{tikzpicture}
	\caption{Hubs should be in the direction of the destination} \label{motivation}
\end{figure}
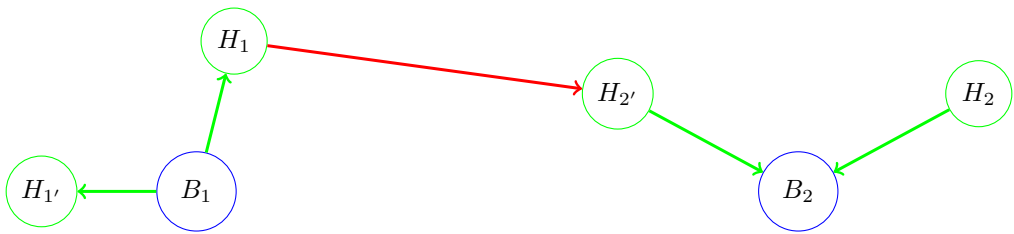

To motivate the idea of the algorithm, consider the task to route one parcel from branch $B_1$ to branch $B_2$ using exactly two of the four possible hubs as in Figure \ref{motivation}. To reduce the problem to facility location, the decision for the first hub (the hub of $B_1$) must be independent of the second hub (the hub of $B_2$). However, ignoring the destination might lead to suboptimal results. In Figure \ref{motivation} hub $H_1$ and $H_{1'}$ are equally far away from $B_1$. Moreover, $H_2$ and $H_{2'}$ are equally far away from $B_2$. Since the task is to get from $B_1$ to $B_2$, using $H_1$ and $H_{2'}$ to reduce the hub-to-hub distance makes sense. These two hubs are especially good since they are in the direction of the destination. Involving the destination for the FLP decision is the contribution of this work and it improves the solution.   

\begin{figure}[H]
	\centering
	\begin{tikzpicture}
		\node[shape=circle,draw=green,minimum size=25pt] (x) at (1.5,2) {$H_1$};
		\node[shape=circle,draw=green,minimum size=25pt] (y) at (11.4,1.3) {$H_2$};
		\node[shape=circle,draw=green,minimum size=25pt] (x2) at (-1.06,0) {$H_{1'}$};
		\node[shape=circle,draw=green,minimum size=25pt] (y2) at (6.6,1.3) {$H_{2'}$};
		\node[shape=circle,draw=blue,minimum size=30pt] (a) at (1,0) {$B_1$};
		\node[shape=circle,draw=blue,minimum size=30pt] (c) at (9,0) {$B_2$};
		\node[shape=circle,draw=red,minimum size=30pt] (m) at (5,0) {$M$};
		
		\path [->,draw=red,line width=0.4mm](x) edge node[left] {} (m);
		\path [->,draw=red,line width=0.4mm](m) edge node[left] {} (y2);
		\path [->,draw=green,line width=0.4mm](a) edge node[left] {} (x);
		\path [->,draw=green,line width=0.4mm](y) edge node[left] {} (c);
		\path [->,draw=green,line width=0.4mm](a) edge node[left] {} (x2);
		\path [->,draw=green,line width=0.4mm](y2) edge node[left] {} (c);
		
	\end{tikzpicture}
	\caption{Subdividing the problem}\label{moti2} 
\end{figure}
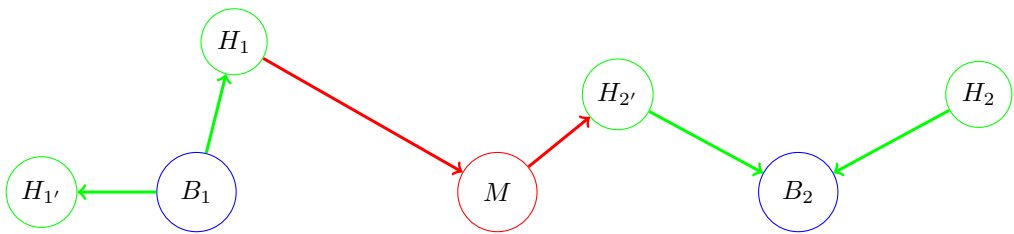

The difficulty is to find good hubs without involving the hub decision of the second hub. The problem can be divided into two parts by adding a mid-point to the problem $M_{B_1,B_2}$, which is halfway between $B_1$ and $B_2$. Since only one tour is considered most of the time, this point is called $M$ for simplicity. One task is to find a short path to the mid-point and one is to find a path from the mid-point to the destination as in Figure \ref{moti2}.   

By the triangle inequality, the tour of Figure \ref{moti2} involving point $M$ can not be shorter than the direct tour as in Figure \ref{motivation}. The main part of the paper is to bound the detour of the second tour. The proposed algorithm using this idea is described as follows:

\pagebreak

\subsection*{Proposed algorithm (PA)}

1. Add for each delivery task $(B_1,B_2)$ a node $M_{B_1,B_2}$ in the HLP instance and define the distances as
\begin{itemize}
	\item $d_{M,M}=0$,
	\item $d_{M_{B_1,B_2},v}=d_{v,M_{B_1,B_2}}=\frac{1}{2}d_{B_1,v}+\frac{1}{2}d_{B_2,v}$ for $v\in V\backslash\{M_{B_1,B_2}\}$.
\end{itemize}

2. Built an FLP instance from the HLP instance in the following manner:

\begin{itemize}
	\item Use the $k$ as $p$ (for the pHM/pHLP) and/or $c$ as facility costs (for the uHLP/pHLP)
	\item Let $\mathcal{C}=\emptyset$ and add for any tour $T_{B_1,B_2}=1$ two cities $\mathcal{C}\leftarrow \mathcal{C} \cup\{ C_{B_1,B_2}\} \cup\{ C_{B_2,B_1}\}$
	\item Use the potential hubs as potential facilities: $\mathcal{F}\leftarrow \mathcal{H}$. To clarify, when talking about facilities or hubs, the facilities will be called $F_1,F_2,...$ and the hubs $H_1,H_2,...$, although $F_i$ directly refers to $H_i$. 
	\item Define the distance function $\Gamma_{i,j}$ as
\end{itemize} 
\begin{flalign*}
	{\Gamma_{(C_{B_1,B_2}),(C_{B_3,B_4})}}:=d_{B_1B_3}+\alpha d_{(M_{B_1,B_2})(M_{B_3,B_4})}& \qquad \text{for branch to branch distances} \\
	{\Gamma_{(C_{B_1,B_2}),F_1}}:=d_{B_1H_1}+\alpha d_{(M_{B_1,B_2})H_1}&\qquad \text{for branch to hub distances} \\
	{\Gamma_{F_1,F_2}}:=d_{H_1H_2}(1+\alpha)& \qquad \text{for hub to hub distances} 
\end{flalign*}

Notice that only the branch-to-hub distance is necessary for the algorithm. However, the other distances need to be defined to show that $\Gamma$ forms a metric.

3. Use a $\gamma$ approximation of a (metric) FLP algorithm

4. Apply this solution to MApHM/MAuFLP/MAHLP by:
\begin{itemize}
	\item Opening a hub if the corresponding facility is opened in the FLP 
	\item Solve the routing optimal
\end{itemize} 

Notice that $c_{B_1,B_2}$ and $c_{B_2,B_1}$ are different vertices and they only have in common that they belong to the same tour and consider the same mid-point in the distance function. Moreover, notice that $c_{B_1,B_2}$ and $c_{B_1,B_3}$ are also different vertices, although both represent a tour starting in $B_1$. If a vector space is given the mid-point can also be set to \[M_{B_1,B_2}=B_1+\frac{1}{2}(B_2-B_1)\] such that $d_{M_{B_1,B_2}v}=d_{vM_{B_1,B_2}}=||B_1+\frac{1}{2}(B_2-B_1)-v||$. This definition directly follows the motivation and works as well. In both cases we directly have 
\begin{align} \label{1gleich2}
d_{B_1M}=\frac{1}{2}d_{B_1B_2}=d_{MB_2}. 
\end{align}

This algorithm produces a valid solution. The pHM and the pHLP open at most $p$ hubs, since the $k$-median problem/ $k$-FLP does so (\eqref{med0},\eqref{hlp0}). Additionally, only open hubs are used for tours since they are opened facilities in the FLP (\eqref{med4},\eqref{med5},\eqref{hlp4}). Furthermore, since metric FLP algorithms are used it is necessary that $\Gamma$ forms a metric. This is shown in the Appendix \ref{1}.

\begin{figure}[H]
	\centering
	\includegraphics[width=0.7\textwidth]{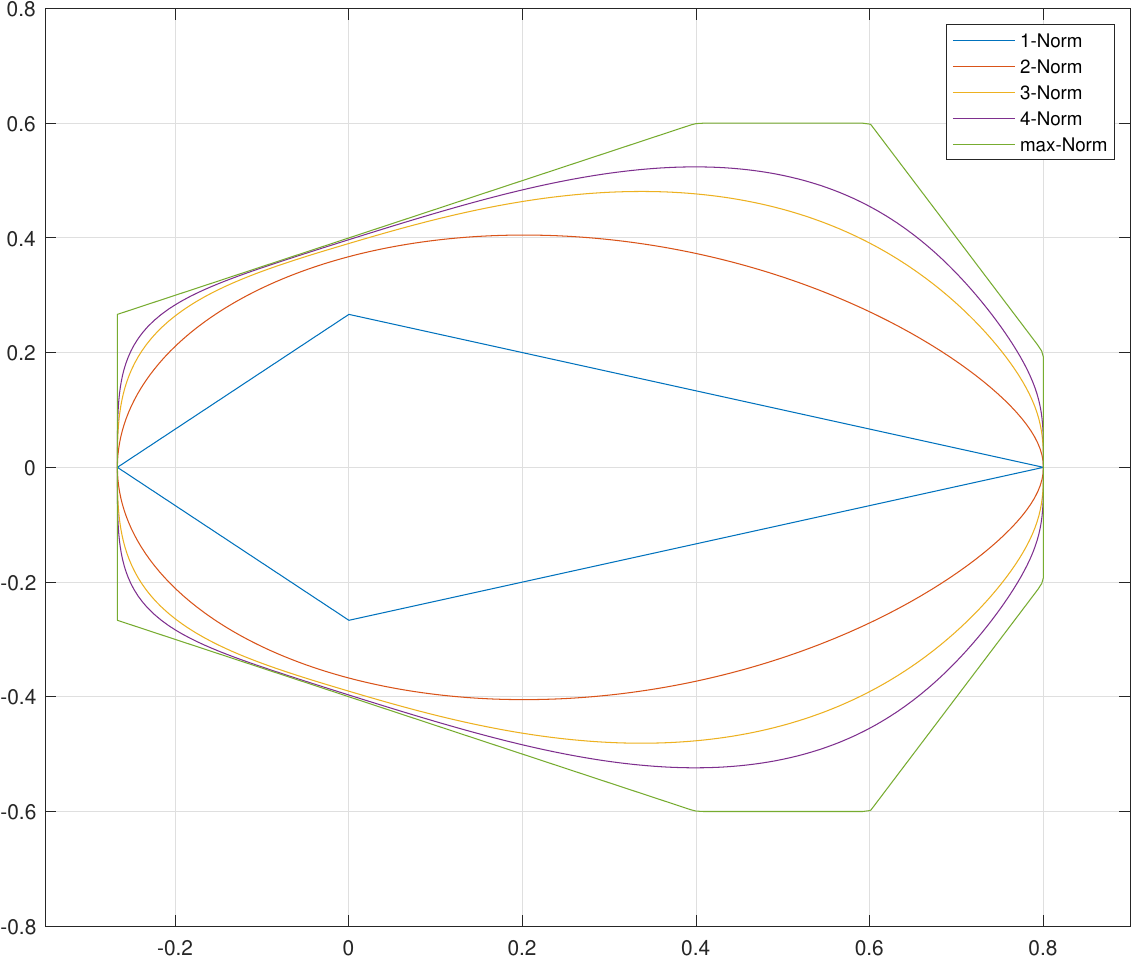}
	\caption{Equally good points in different metrics} \label{aufkommen}
\end{figure}

To visualize the algorithm's idea, consider two branches $B_1$ and $B_2$ with coordinates $(0,0)$ and $(2,0)$ and $\alpha=0.5$ in a vector space. Hence, $M=(1,0)$. A potential hub $H_1=(x,y)$ on the way from $B_1$ to $M$ has costs $\sqrt[p]{|x|^p+|y|^p}+\alpha\sqrt[p]{|1-x|^p+|y|^p}$ in a $p$-norm. In Figure \ref{aufkommen}, any point having exactly cost $0.9$ is shown for different norms.

This figure illustrates that the proposed cost function considers both intuitions. On the one hand, close hubs are preferred over far away hubs; on the other hand, hubs on the way to the destination $B_2$ are preferred over hubs in other directions. For instance, the hubs $H'=(-0.2\bar{6},0)$ and $H''=(0.8,0)$ have both costs of $0.9$, meaning they are considered equally good first hubs. $H'$ has the advantage of being close to $B_1$ and $H''$ has the advantage of being in the destination's direction. If $\alpha$ is reduced, it is more important to reduce the branch-to-hub distance; if $\alpha$ is enlarged, it is more important to reduce the hub-to-hub distance by preferring hubs close to the mid-point. For $\alpha=0$, this graph would be the (scaled) unit circle for the different norms, which makes sense since the hub-to-hub distance can be neglected. 

In the next section, the algorithm is bounded.

\section{Approximation guarantee}\label{nor}

To obtain the claimed bound, the proposed algorithm (PA) as well as the algorithm of \cite{benedito2019approximation} (BaP) will be used. The solution with the smaller costs will than be chosen. The PA obtain better results than BaP for smaller values of $\alpha$ and BaP obtains in theory better results for larger values of $\alpha$.

\begin{theo}
	Applying the PA and BaP with the $k$-median algorithm of \cite{byrka2017improved} is a $3.451$ approximation algorithm for the MApHM. For the MAuHLP applying the algorithm of \cite{li20131} for uFLP gives a $2.173$ approximation algorithm. Applying the algorithm of \cite{zhang2007new} is a $4.552$ approximation algorithm for the MApHLP.
\end{theo}

In any proof, the bounds are shown for one fixed delivery task. Since the proofs hold for any delivery task, the bound also holds for the whole solution.

Proof: Since the PA considers an optimal routing in step four, any bound for specific routing strategies hold simultaneously. Therefore, two routing strategies similarly to BaP are considered. Let $B_1,B_2 \in \mathcal{B}$ and $H_1,H_2\in \mathcal{H}$ such that $C_{B_1,B_2}$ is connected to facility $F_1$ and $C_{B_2,B_1}$ to $F_2$. In other words, for the routing from $B_1$ to $B_2$ the facility location algorithm connected the corresponding cities to $F_1$ respectively $F_2$.   

\textbf{\textcolor{red}{Routing strategy 1:}} A routing is performed using the hubs corresponding to the city-facility connections. In other words $B_1\rightarrow H_1 \rightarrow H_2 \rightarrow B_2$ is used. 

\textbf{\textcolor{brown}{Routing strategy 2:}} Only one hub is used which is connected to one of the two branches by a city-facility connection. This hub is than connected to both branches. In other words $B_1\rightarrow H_1 \rightarrow B_2$ or $B_1\rightarrow H_2 \rightarrow B_2$ is used.

\begin{figure}[H]
	\centering
	\begin{tikzpicture}
	\node[shape=circle,draw=green,minimum size=25pt] (x) at (1.5,2) {$H_1$};
	\node[shape=circle,draw=green,minimum size=25pt] (y2) at (6.6,1.3) {$H_{2}$};
	\node[shape=circle,draw=blue,minimum size=30pt] (a) at (1,0) {$B_1$};
	\node[shape=circle,draw=blue,minimum size=30pt] (c) at (9,0) {$B_2$};
	
	\path [->,draw=red,line width=0.4mm](x) edge node[left] {} (y2);
	\path [->,draw=red,bend left, line width=0.4mm](a) edge node[left] {} (x);
	\path [->,draw=red,line width=0.4mm](y2) edge node[left] {} (c);
	
	\path [->,draw=brown,bend right,line width=0.4mm](a) edge node[left] {} (x);
	\path [->,draw=brown,line width=0.4mm](x) edge node[left] {} (c);
	
	\end{tikzpicture}
	\caption{Routing strategies} \label{strategies}
\end{figure}
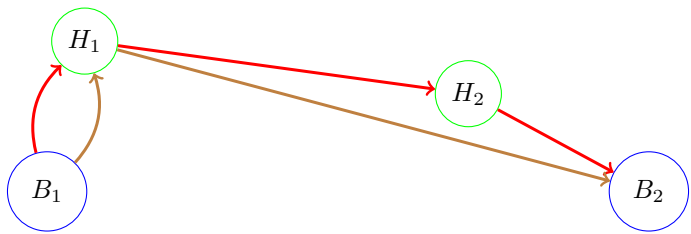

Figure \ref{strategies} visualize both strategies. Strategy one is especially good for low values of $\alpha$ since the hub to hub connection is discounted. Strategy two is good for $\alpha$ close to $1$. Notice that for $\alpha=1$ by the triangle inequality strategy $1$ can not outperform strategy $2$.   

In the following for both strategies an approximation bound is established.

\begin{lem} \label{lem0}
	The proposed algorithm with routing strategy 1 is a $(1+\alpha)\gamma$ approximation algorithm.
\end{lem}

\begin{lem}\label{lem3}
 	The proposed algorithm with routing strategy 2 is a $\gamma+\frac{1}{(1+\alpha)\alpha}$ approximation algorithm.
\end{lem}

In BaP, the idea of using the presented routing strategies is used as well, reaching the guarantees $ \frac{1}{\alpha}$ and $\gamma+\alpha(1+\gamma)$ respectively yielding together to a $1+\gamma$ bound. In Table \ref{difalgos} the different bounds are shown.
\begin{table}[!h]
	\begin{center}
		\begin{tabular}[h]{c|c|c}
			Idea  &   BaP & PA  \\
			\hline
			&&\\
			Using one hub & BaP$_1=\frac{1}{\alpha}$& PA$_1=\gamma+\frac{1}{(1+\alpha)\alpha}$ \\
			&&\\
			Using two hubs & BaP$_2=(1+\alpha)\gamma+\alpha$  & PA$_2=(1+\alpha)\gamma $ \\
			&&\\
		\end{tabular}
		\caption{Proven bounds}  \label{difalgos}
	\end{center}
\end{table}

Notice that BaP$_1\leq$ PA$_1$ and PA$_2\leq$ BAP$_2$. Hence, if both algorithms are used and the best solution is applied, the guarantee can be decreased to \[\min\left( (1+\alpha)\gamma,\frac{1}{\alpha}\right). \] In Figure \ref{figure} any bound is visualized using $\gamma=2.675$, which is the $k$-median guarantee of the algorithm of \cite{byrka2017improved} (for MApHM). Using BaP would guarantee an approximation bound of the intersection between the black and blue function ($3.675$), which has been the best approximation bound before. Running the proposed algorithm guarantees the approximation bound of the intersection between the orange and red lines ($4.011$). Running both algorithms and taking the best result guarantees any bound. Therefore, the black and red lines' intersection ($3.451$) can be obtained with this method.

\begin{figure}[H]	
	\begin{minipage}{.5\textwidth}
	\begin{center}
		\begin{tikzpicture}
		\centering
		\begin{axis}[
		domain=0:1,
		xmin=0, xmax=1,
		ymin=2, ymax=6,
		samples=400,
		axis y line=center,
		axis x line=middle,
		xlabel={$\alpha$},
		ylabel={$\gamma'$},
		]
		
		\addplot+[mark=none,color=black] {1/x};
		\addplot+[mark=none,color=blue] {2.675+x*(3.675)};
		\addplot+[mark=none,color=orange] {2.675+(1/((1+x)*x))};
		\addplot+[mark=none,color=red] {2.675*(1+x)};
		\legend{BAP$_1$,BAP$_2$,PA$_1$,PA$_2$}	
		\end{axis}
		\end{tikzpicture}
	\end{center}
\caption{MApHM: $\gamma=2.675$} \label{figure}
	\end{minipage}%
\begin{minipage}{.5\textwidth}
	\begin{center}
		\begin{tikzpicture}
		\centering
		\begin{axis}[
		domain=0:1,
		xmin=0, xmax=1,
		ymin=1, ymax=5,
		samples=400,
		axis y line=center,
		axis x line=middle,
		xlabel={$\alpha$},
		ylabel={$\gamma'$},
		]
		
		\addplot+[mark=none,color=black] {1/x};
		\addplot+[mark=none,color=blue] {1.488+x*(2.488)};
		\addplot+[mark=none,color=orange] {1.488+(1/((1+x)*x))};
		\addplot+[mark=none,color=red] {1.488*(1+x)};
		\legend{BAP$_1$,BAP$_2$,PA$_1$,PA$_2$}	
		\end{axis}
		\end{tikzpicture}
	\end{center}
	\caption{MAuHLP: $1.488$} \label{figure2}
\end{minipage}%
\end{figure}

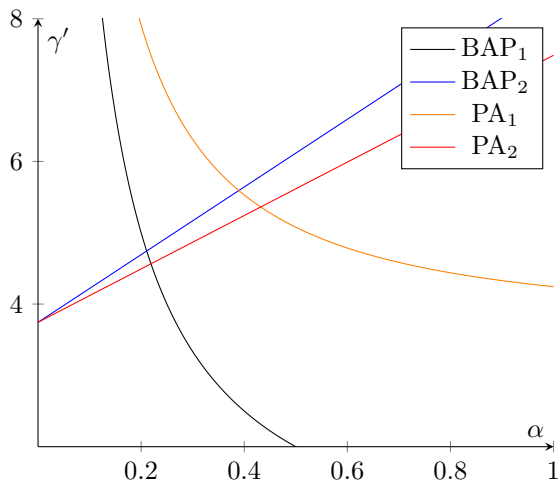
\begin{figure}[H]
	\begin{center}
		\begin{tikzpicture}
		\centering
		\begin{axis}[
		domain=0:1,
		xmin=0, xmax=1,
		ymin=2, ymax=8,
		samples=400,
		axis y line=center,
		axis x line=middle,
		xlabel={$\alpha$},
		ylabel={$\gamma'$},
		]
		
		\addplot+[mark=none,color=black] {1/x};
		\addplot+[mark=none,color=blue] {3.744+x*(4.744)};
		\addplot+[mark=none,color=orange] {3.744+(1/((1+x)*x))};
		\addplot+[mark=none,color=red] {3.744*(1+x)};
		\legend{BAP$_1$,BAP$_2$,PA$_1$,PA$_2$}	
		\end{axis}
		\end{tikzpicture}
	\end{center}
	\caption{MApHLP: $\gamma=2+\sqrt{3}$} \label{figure3}
\end{figure}
For the MAuHLP this is analogously using the $\gamma=1.488$ bound by \cite{li20131} as visualized in Figure \ref{figure2}. In Figure \ref{figure3} this is visualized with the $\gamma=2+\sqrt{3}$ bound of \cite{zhang2007new} for the MApFLP. In Table \ref{result} the decreased bounds are presented.

\begin{table}[!h]
	\begin{center}
		\begin{tabular}[h]{c|c|c|c|c|c}
			Problem  & Reduced to & Used factor &  BaP & PA &combination  \\
			\hline
			MApHM & $k$-median & $2.675$\cite{byrka2017improved} & $3.675$ &$4.011$& $3.451$ \\
			MAuHLP & FLP & $1.488$ \cite{li20131}& $2.488$ & $2.441$ &$2.173 $ \\
			MApHLP & $k$-FLP & $2+\sqrt{3}+\epsilon$ \cite{zhang2007new} & $4.733$ &$5.346$& $4.552 $ \\
		\end{tabular}
		\caption{Decreased bounds}  \label{result}
	\end{center}
\end{table}

As a result, the approximation bound of the MApHM is improved to $3.451$, the approximation bound of the MAuHLP is improved to $2.173 $ and the bound of the MApHLP is improved to $4.552 $ . \qed
\section{Proofs}\label{proof}
In this section Lemma \ref{lem0} and \ref{lem3} are shown. Let $\text{OPT}_{HLP}$ be the optimal objective value of the HLP instance, $\text{OPT}_{FLP}$ the of the FLP instance and $ALG_{HLP}$, $ALG_{FLP}$ the corresponding objective values of the algorithms solution. 

The lemmata are shown by using the additional following lemma:

\begin{lem}\label{lem5}
	It holds $\text{ALG}_{FLP}\leq \gamma(1+\alpha) OPT_{HLP}$.
\end{lem}

\subsection{Proof of Lemma \ref{lem0}}

\textit{The proposed algorithm with routing strategy 1 is a $(1+\alpha)\gamma$ approximation algorithm.}

Proof: In strategy 1 as in Figure \ref{strategies} the routing of the solution is done according to the facility location connections. For any FLP solution $B_1\rightarrow H_1$ and $B_2\rightarrow H_2$ with mid-point $M$ holds

\[ALG_{FLP}=d_{B_1H_1}+\alpha d_{H_1M}+\alpha d_{MH_2}+d_{ H_2B_2}  \] 
\[ \geq_\eqref{dreieck} d_{B_1H_1}+\alpha d_{H_1H_2}+d_{H_2B_2}=ALG_{HLP}. \]
The set-up costs were neglected since they are equal for $ALG_{FLP}$ and $ALG_{HLP}$ by the definition of the strategy.
Together with Lemma \ref{lem5} directly \[ALG_{HLP}\leq (1+\alpha)\gamma OPT_{HLP}  \] follows.\qed

\subsection{Proof of Lemma \ref{lem3}}

\textit{The proposed algorithm with routing strategy 2 is a $\gamma+\frac{1}{(1+\alpha)\alpha}$ approximation algorithm.}

Proof: In strategy 2 as in Figure \ref{strategies} only one hub is used in the HLP routing. 
As before, the set-up costs for the MAuHLP can be neglected since they are the same in $ALG_{FLP}$ and $ALG_{HLP}$.

W.l.o.g. let \begin{align} d_{B_1H_1}+\alpha d_{H_1M}\leq d_{B_2H_2}+\alpha d_{MH_2}\label{formula}. \end{align}

According to strategy 2 the routing is $B_1\rightarrow H_1\rightarrow B_2$.
This gives:
\[ ALG_{HLP}\leq d_{B_1H_1}+d_{H_1B_2} \] 
\[ \leq_\eqref{dreieck} d_{B_1H_1}+d_{H_1M}+d_{MB_2}\] 
\[ \leq_\eqref{dreieck} d_{B_1H_1}+\frac{2\alpha}{1+\alpha}d_{H_1M}+\frac{1-\alpha}{1+\alpha}(d_{B_1M}+d_{B_1H_1})+d_{MB_2} \] 
\[ = \frac{2}{1+\alpha}(d_{B_1H_1}+\alpha d_{H_1M})+\frac{1-\alpha}{1+\alpha} \cdot d_{B_1M}+d_{MB_2}\] 
\[\leq_\eqref{formula}\frac{2}{1+\alpha}\cdot\frac{1}{2}(d_{B_1H_1}+\alpha d_{H_1M}+d_{B_2H_2}+\alpha d_{MH_2})+\frac{1-\alpha}{1+\alpha} \cdot d_{B_1M}+d_{MB_2} \]
\[=\frac{1}{1+\alpha}ALG_{FLP}+\frac{1-\alpha}{1+\alpha} \cdot d_{B_1M}+d_{MB_2} \]
\[=_\eqref{1gleich2}\frac{1}{1+\alpha}ALG_{FLP}+ \frac{2}{1+\alpha} \cdot d_{B_1B_2} \cdot \frac{1}{2}\]
\[ \leq_{\ref{lem5}} \frac{1}{1+\alpha}(1+\alpha)\gamma \text{OPT}_{\text{HLP}}+  \frac{1}{1+\alpha}  \cdot d_{B_1B_2}\]
\[ \leq\left(  \gamma+ \frac{1}{1+\alpha}\cdot \frac{1}{\alpha} \right) \cdot \text{OPT}_{\text{HLP}}\]
\[ =\left(  \gamma+\frac{1}{(1+\alpha)\alpha} \right) \cdot \text{OPT}_{\text{HLP}}\]
In the last inequality it is used that any solution has at least cost of the discounted direct connection between $B_1$ and $B_2$ such that $\text{OPT}_{\text{HLP}}\geq \alpha\cdot d_{B_1B_2}$. \qed

\subsection{Proof of Lemma \ref{lem5}}

\textit{It holds $\text{ALG}_{FLP}\leq \gamma(1+\alpha) OPT_{HLP}$.}

Proof: Since a $\gamma$-approximation is used $ ALG_{FLP}\leq \gamma \text{OPT}_{FLP}$ holds. 

It is left to show that $OPT_{FLP}\leq (1+\alpha)\cdot OPT_{HLP}$. Let an optimal HLP solution be given. An FLP solution $I$ can be constructed by using the connections according to the HLP solution. The set-up costs can be neglected since they are the same in $I$ and $OPT_{HLP}$. Since $I$ is a valid FLP solution $OPT_{FLP}\leq cost(I)$. Again fix a tour from $B_1$ via $H_1$ and $H_2$ to $B_2$ in the optimal hub location solution.

\[ {OPT}_{FLP}\leq cost(I)\]
\[=d_{B_1H_1}+\alpha d_{H_1M}+\alpha d_{MH_2}+ d_{H_2B_2}  \]
\[=d_{B_1H_1}+\alpha \left( \frac{1}{2}d_{B_1H_1}+\frac{1}{2}d_{B_2H_1}\right) +\alpha \left( \frac{1}{2}d_{B_1H_2}+\frac{1}{2}d_{B_2H_2}\right)+ d_{H_2B_2}\]
\[=d_{B_1H_1}+\alpha \left( d_{B_1H_1}+\frac{1}{2}\left( d_{B_2H_1} -d_{H_2B_2} \right)+ \frac{1}{2}\left( d_{B_1H_2}-d_{B_1H_1}\right) +d_{B_2H_2}\right)+ d_{H_2B_2}\]
\[\leq_\eqref{dreieck}d_{B_1H_1}+\alpha \left( d_{B_1H_1}+\frac{1}{2}\left( d_{H_1H_2} \right)+ \frac{1}{2}\left( d_{H_1H_2}\right) +d_{B_2H_2}\right)+ d_{H_2B_2}\]
\[\leq (1+\alpha) \left( d_{B_1H_1}+\alpha d_{H_1H_2}+d_{H_2B_2}\right)  \]
\[=(1+\alpha)\cdot OPT_{HLP} \]

The proof for the alternative vector space definition of $M$ is in the appendix at \ref{dipdidu}.\qed

\section{Computational results} \label{prac}

In this section, the computational results of the proposed algorithm are shown. The described algorithm is compared to BaP. 

The algorithms differ only by the definition of the distance between a branch and hub in step $2$ of the algorithm. Instead of defining \[{\Gamma_{(C_{B_1,B_2}),F_1}}:=d_{B_1H_1}+\alpha d_{(M_{B_1,B_2})H_1}\] for branch to hub distances, BaP defines the distance as \[{\Gamma_{(C_{B_1,B_2}),F_1}}:=d_{B_1H_1}.\] 

For both algorithms an optimal routing and the same FLP algorithms are applied. A simple greedy algorithm for the $k$-median problem was used to obtain reasonable results for huge instances quickly. The algorithm starts with an empty set of facilities and in each of the $k$ steps, it adds the facility, which reduces the maximal costs for the FLP in this iteration. 

Similarly, for the uncapacitated FLP, the algorithm of \cite{hochbaum1982heuristics} is used, which defines all cities as uncovered in the beginning. In each iteration, it greedily covers a set of uncovered cities $\tilde{C}$ by a facility $f$ minimizing $\frac{c_f+\sum_{c\in\tilde{C}}\Gamma_{f,c}}{|\tilde{C}|}$.  

Benchmark instances exist at \cite{campbell1994integer} or from the Australian post at \cite{ernst1996efficient}. However, only a few instances were considered, and each instance is small. Hence, new instances were created to ensure enough test cases.  

Three sets of test instances were created:

Small-sized instances with $1,000$ delivery tasks, $50$ branches, $100$ hubs and $1,000$ samples.

Medium-sized instances with $5,000$ delivery tasks, $100$ branches, $200$ hubs and $200$ samples.

Big instances with $20,000$ delivery tasks, $1,000$ branches, $400$ hubs and $100$ samples. 

All instances were too huge to get results from an optimal solver in a reasonable time. 

The locations were drawn uniformly for two dimensions in $[0,1]$. For MAuHLP, the first \linebreak$100$ small-sized instances were considered. Additionally, the set-up costs were either uniformly \linebreak drawn from $[0,1.2]$ or set to $1$. Any run's test instances and objectives can be obtained at \linebreak{\url{http://dx.doi.org/10.17877/DE290R-23200}.} Additionally, in the instances volumes were given, which were neglected for these problem statements. An extension would be to weight tours differently.

Table \ref{tabel2} shows median values for the $2$-norm MApHM.

\begin{table}[!h]
	\begin{center}
		\begin{tabular}[h]{c|c|c||c|c}
			Instance  & $p$ & $\alpha$ &  BaP & PA  \\
			\hline
			small & $6$ & $0.4$ & $457.45$ & $453.51$ \\
			small & $6$ & $0.8$ & $559.51$ &  $552.77$ \\
			medium & $10$ & $0.4$ & $2022.89$ & $2008.86$ \\
			medium & $10$ & $0.8$ & $2646.60$ &  $2613.79$ \\
			medium & $2$ & $0.4$ & $3348.34$ & $3343.79$ \\
			medium & $2$ & $0.8$ & $3455.82$ &  $3446.42$ \\
			medium & $20$ & $0.4$ & $1700.68$ & $1684.64$ \\
			medium & $20$ & $0.8$ & $2472.76$ &  $2443.30$ \\
			medium\tablefootnote{In this instance a $20$ norm was considered instead of a $2$ norm.\label{footnote}} & $10$ & $0.4$ & $1792.72$ & $1780.30$ \\
			medium$^{\ref{footnote}}$& $10$ & $0.8$ & $2331.59$ &  $2300.02$ \\
			big & $12$ & $0.4$ & $8115.80$ & $8040.45$ \\
			big & $12$ & $0.8$ & $10591.53$ &  $10444.19$ \\
			
		\end{tabular}
		\caption{Average objective value for MApHM}  \label{tabel2}
	\end{center}
\end{table}

In any test case, the proposed algorithm significantly improves the result concerning BaP.

The results for MAuHLP are presented in Table \ref{tabel3}.

\begin{table}[!h]
	\begin{center}
		\begin{tabular}[h]{c|c|c||c|c}
			Instance  &  set up &$\alpha$&  BaP & PA  \\
			\hline
			small& uniform & $0.4$ & $330.45$ &$329.88$\\
			small& uniform & $0.8$ & $504.58$ & $499.48$\\
			small& set1 & $0.4$ &$342.22$ & $341.70$\\
			small& set1 & $0.8$ & $516.84$ & $516.75$\\
		
			small$^{\ref{footnote}}$& uniform & $0.4$ & $296.45$ &$295.64$\\
			small$^{\ref{footnote}}$& uniform & $0.8$ & $448.33$ & $443.49$\\
			small$^{\ref{footnote}}$& set1 & $0.4$ &$308.31$ & $307.94$\\
			small$^{\ref{footnote}}$& set1 & $0.8$ & $460.49$ & $461.11$\\			
			
			small\tablefootnote{In this instances any set-up costs were doubled.\label{footnote2}}& uniform & $0.4$ & $351.05$ &$349.95$\\
			small$^{\ref{footnote2}}$& uniform & $0.8$ & $523.79$ & $515.64$\\
			small$^{\ref{footnote2}}$& set1 & $0.4$ &$373.37$ & $373.37$\\
			small$^{\ref{footnote2}}$& set1 & $0.8$ & $547.30$ & $547.18$\\

		\end{tabular}
		\caption{Average objective value for MAuHLP}  \label{tabel3}
	\end{center}
\end{table}

For the MAuHLP, the PA outperforms BaP in any but one test instance. Moreover, the tests suggest that the difference between both algorithms is for the MApHM more significant. For the MApHLP the objective values depend on the relation between $k$ and the set-up costs. For large $k$ the MAuHLP solutions were received and for small $k$ results similar to the MApHM solutions. 

\pagebreak

In any but one test case, PA outperforms BaP such that beyond the theoretical improvement, the tests indicate that the proposed algorithm is, from a practical point of view, superior and should be used when the instance size is too large or the time bound is too small for exact algorithms.

\section*{Acknowledgments}

Special thanks to Anna Schroeter, Dorothee Henke and Nele Pommerening for the helpful discussions. Furthermore, thanks to Aleksandra "Ola" Grochala for helping with the implementation. 

\pagebreak
 \bibliographystyle{agsm}
 \bibliography{bib} 

\pagebreak
\section*{Appendix}
\appendix
\section{$\Gamma$ forms a metric} \label{1}

Metric FLP algorithms can only be used if the created FLP instance defines a metric. Hence, it needs to shown that $\Gamma$ forms a metric for the FLPs.

As described, $\Gamma$ refers to the distances in the created graph for metric FLPs and $d$ refers to the distance of the input graph. In the following, let $C_{B_1,B_2},C_{B_3,B_4},C_{B_5,B_6}\in \mathcal{C}$ be any city in the FLP with $B_1,B_2,B_3,B_4,B_5,B_6\in \mathcal{B}$ are the corresponding branches and $M_{B_1,B_2},M_{B_3,B_4},M_{B_5,B_6}$ are the corresponding mid-points of the tours in the HLP. In addition, let $F_1,F_2,F_3\in \mathcal{F}$ be potential facilities in the FLP. Definite\eqref{ident}, symmetry\eqref{sym} and the triangle inequality\eqref{dreieck} must be proven for each combination of cities and facilities as defined in the algorithm.  

1. Definite holds due to \[ \Gamma_{(C_{B_1,B_2}),(C_{B_1,B_2})}=d_{B_1B_1}+\alpha d_{(M_{B_1,B_2})(M_{B_1,B_2})}=0 \]
and \[ \Gamma_{F_1,F_1}=d_{H_1H_1}(1+\alpha)=0\]

2. Symmetry directly follows from the definition.

3. The triangle inequality will be shown for any case. We distinguish the cases \textbf{1} between two cities, \textbf{2} between one city and one facility and \textbf{3} between two facilities. Each case can be further distinguished if the shortcut is done through a city \textbf{.1} or a facility \textbf{.2}. Any equation uses that $ d$ forms a metric such that the triangle inequality on $d$ can be used.

\textbf{1.1:} \[\Gamma_{(C_{B_1,B_2}),(C_{B_3,B_4})}=d_{B_1B_3}+\alpha d_{(M_{B_1,B_2})(M_{B_3,B_4})}\]\[\leq	 d_{B_1B_5}+d_{B_5B_3}+\alpha d_{(M_{B_1,B_2})(M_{B_5,B_6})}+\alpha d_{(M_{B_5,B_6})(M_{B_3,B_4})} \]\[=\Gamma_{(C_{B_1,B_2}),(C_{B_5,B_6})}+\Gamma_{(C_{B_5,B_6}),(C_{B_3,B_4})}\]

\textbf{1.2:} \[\Gamma_{(C_{B_1,B_2}),(C_{B_3,B_4})}=d_{B_1B_3}+\alpha d_{(M_{B_1,B_2})(M_{B_3,B_4})}\]\[\leq d_{B_1H_1}+d_{H_1B_3}+\alpha d_{(M_{B_1,B_2})H_1}+\alpha d_{H_1(M_{B_3,B_4})} \]\[=\Gamma_{(C_{B_1,B_2}),F_1}+\Gamma_{F_1,(C_{B_3,B_4})}\]

\textbf{2.1:} \[\Gamma_{(C_{B_1,B_2}),F_1}=d_{B_1H_1}+\alpha d_{(M_{B_1,B_2})H_1}\]\[\leq d_{B_1B_3}+d_{B_3H_1}+\alpha d_{(M_{B_1,B_2})(M_{B_3,B_4})}+\alpha d_{(M_{B_3,B_4})H_1} \]\[=\Gamma_{(C_{B_1,B_2}),(C_{B_3,B_4})}+\Gamma_{(C_{B_3,B_4}),F_1}\]

\textbf{2.2:} \[\Gamma_{(C_{B_1,B_2}),F_1}=d_{B_1H_1}+\alpha d_{(M_{B_1,B_2})H_1}\]\[\leq d_{B_1H_2}+d_{H_2H_1}+\alpha d_{(M_{B_1,B_2})H_2}+\alpha d_{H_2H_1}\]\[=\Gamma_{(C_{B_1,B_2}),F_2}+\Gamma_{F_2,F_1}\]

\textbf{3.1:} \[\Gamma_{F_1,F_2}=d_{H_1,H_2}(1+\alpha)\]\[\leq d_{H_1B_1}+d_{B_1H_2}+\alpha d_{H_1(M_{B_1,B_2})}+\alpha d_{(M_{B_1,B_2})H_2}\]\[=\Gamma_{F_1,(C_{B_1,B_2})}+\Gamma_{(C_{B_1,B_2}),F_2}\]

\textbf{3.2:} \[\Gamma_{F_1,F_2}=d_{H_1H_2}(1+\alpha)\]\[\leq (d_{H_1H_3}+d_{H_3H_2})(1+\alpha) \]\[=\Gamma_{F_1,F_3}+\Gamma_{F_3,F_2}\]

This proves that $\Gamma$ forms a metric.\qed

\section{Prove of lemma \ref{lem5} for alternative definition of $M$}\label{dipdidu}

W.l.o.g let $M$ be in the origin, such that $||M||=0$. Furthermore, let $||H_1||\geq ||H_2||$. Since, $M$ is the mid-point $||B_1||=||-B_2 ||$. Than 
\[ ALG_{HLP}\leq ||B_1-H_1||+||H_1-B_2|| \] 
\[ \leq_\eqref{dreieck} ||B_1-H_1||+||H_1||+||B_2||\] 
\[ \leq_\eqref{dreieck} ||B_1-H_1||+\frac{2\alpha}{1+\alpha}||H_1||+\frac{1-\alpha}{1+\alpha}(||B_1||+||B_1-H_1||)+||B_2|| \] 
\[ = \frac{2}{1+\alpha}(||B_1-H_1||+\alpha||H_1||)+\frac{1-\alpha}{1+\alpha} \cdot||B_1||+||B_2||\] 
\[\leq_\eqref{formula}\frac{2}{1+\alpha}\cdot\frac{1}{2}(||B_1-H_1||+\alpha||H_1||+||B_2-H_2||+\alpha||H_2||)+\frac{1-\alpha}{1+\alpha} \cdot||B_1||+||B_2|| \]
\[=\frac{1}{1+\alpha}ALG_{FLP}+\frac{1-\alpha}{1+\alpha} \cdot||B_1||+||B_2|| \]
\[=\frac{1}{1+\alpha}ALG_{FLP}+ \frac{2}{1+\alpha} \cdot ||B_1|| \]
\[ \leq_{\ref{lem5}} \frac{1}{1+\alpha}(1+\alpha)\gamma \text{OPT}_{\text{HLP}}+  \frac{2}{1+\alpha}  \cdot ||B_1||\]
\[ \leq\left(  \gamma+ \frac{2}{1+\alpha}\cdot \frac{1}{2\alpha} \right) \cdot \text{OPT}_{\text{HLP}}\]
\[ =\left(  \gamma+\frac{1}{(1+\alpha)\alpha} \right) \cdot \text{OPT}_{\text{HLP}}\]\qed

\end{document}